\documentclass{ws-procs9x6}

\newcommand{\beq}{\begin{equation}}
\newcommand{\eeq}{\end{equation}}
\newcommand{\beqa}{\begin{eqnarray}}
\newcommand{\eeqa}{\end{eqnarray}}

\newcommand{\lap}{\lower.5ex\hbox{$\; \buildrel < \over \sim \;$}}
\newcommand{\gap}{\lower.5ex\hbox{$\; \buildrel > \over \sim \;$}}

\begin{document}

\title{PROBING GENERAL RELATIVITY ON \\ THE SCALES OF COSMOLOGY}

\author{P. J. E. PEEBLES}
\address{Joseph Henry Laboratories,
Princeton University, 
\\ Princeton NJ 08544, USA
\\ E-mail: pjep@princeton.edu}  

\maketitle

\abstracts{
The cosmological tests are tight enough now to show that the Friedmann-Lema\^\i tre $\Lambda$CDM cosmological model almost certainly is a useful approximation. This means general relativity theory passes significant tests of the extrapolation of some fifteen orders of magnitude from the length scales of the precision tests of gravity physics.}

\section{Introduction}

Cosmology is tested by checking its consistency with  observational constraints that are substantially more numerous than the freedom to adjust the theory and the interpretations of the observations. We are reaching this goal for some aspects of cosmology: particularly impressive is the abundance of evidence that the matter density parameter is in the range 
\beq
0.15\lap\Omega _m\lap 0.3. \label{eq:omegam}
\eeq
This certainly is not precision cosmology, but the consistency of estimates that depend on many different aspects of the theory and a broad variety of astronomical methods show that it almost certainly is accurate cosmology, that is, that we have not been misled by systematic errors. That means general relativity theory passes significant empirical tests on the scale of the observable universe. The measurements supporting this remarkable advance are reviewed in Sec.~3. After brief comments on the more modest network of evidence for the detection of $\Lambda$, in Sec.~4, I conclude with a qualitative assessment of our understanding of the physics of spacetime and gravity on the scales of cosmology, and the case for developing a more systematic framework for judging the empirical evidence. 

\section{The Standard Cosmology}

It will be useful for what follows to consider the elements of the present standard Friedmann-Lema\^\i tre $\Lambda$CDM cosmology. In this model, 
\begin{enumerate}
\item gravity and spacetime are described by general relativity theory; 
\item the stress-energy tensor represents 
\begin{enumerate}
\item textbook physics in the visible sector, with baryon density $\Omega _{\rm b}=0.04$ and thermal radiation at $T_o=2.725$~K;    
\item cold dark matter, with $\Omega _{\rm CDM} + \Omega _{\rm b}$ in the range of Eq.~(\ref{eq:omegam}), and dark energy sufficient for flat space sections; 
\end{enumerate}
\item initial conditions that describe 
\begin{enumerate}
\item expansion from very high redshift;
\item growth of structure out of adiabatic, Gaussian, and scale-invariant departures from homogeneity and isotropy. 
\end{enumerate}
\end{enumerate}
It is a measure of the durability of these concepts --- and the slow progress of research in cosmology --- that much of this picture was under discussion in the 1930s. Eddington disliked the idea that our universe expanded from a very different state, but Lema\^\i tre\cite{Lem31} celebrated it as a chance to learn new physics. Tolman showed that homogeneous and isotropic expansion cools a uniform sea of blackbody radiation while preserving the thermal spectrum. I know no evidence that he thought there might be an observable consequence; Gamow introduced that in the 1940s, with his proposal for the formation of deuterium and heavier elements. In the 1930s Zwicky and Hubble knew that the masses of nearby clusters of galaxies appear to be larger than the mass in the luminous parts of the galaxies, but I suspect it would have taken considerable  explanation to convince them that Gamow's theory of the origin of the light elements requires that most of the missing mass --- or  dark matter --- has to be nonbaryonic. The other component of the dark sector would have been easy to explain in the 1930s, and perhaps also the reason for its new name, because Lema\^\i tre, Eddington, and others were aware of the interpretation of $\Lambda$ as an energy density with a curious equation of state (as reviewed in Sec.~III.B.3 in Ref.~\refcite{PR}). Pauli knew the problem with the quantum vacuum energy density, though perhaps not the possible relation to the issue of $\Lambda$. In his 1933 article\cite{Pauli1933} on quantum mechanics Pauli advised ignoring the zero-point energy of the electromagnetic field, because its energy density is absurd, but I expect he would have agreed that his prescription makes no sense. This is a good description of the present understanding of the vacuum energy. 

In the 1930s Lema\^\i tre proposed that gravity drove structure formation. Four decades ago this was still debated. Two decades ago the gravitational instability picture was generally accepted but initial conditions were debated, whether set by explosions or cosmic strings or global monopoles, or to be stipulated as primeval isocurvature or adiabatic departures from homogeneity. The community has fixed on a variant of the last, in 3b, because it fits the observations. That is good science, but to be borne in mind when we consider the freedom to adjust the theory to fit the observations. 

Einstein and de~Sitter model proposed their model in 1932. The evidence that the mass density is lower than the Einstein-de~Sitter value would not have surprised de~Sitter. Would the evidence that the low mass density is accompanied by $\Lambda$ rather than space curvature have disturbed or pleased Einstein? It was his invention, after all.

\begin{table}[ph]
\tbl{Constraints on the matter density parameter.\vspace*{1pt}}
{\footnotesize
\begin{tabular}{|rlcc|}
\hline
{} & {} &{} &{} \\[-1.5ex]
{} &Measurement & Scale & $\Omega _m$ \\[1ex]
\hline
{} &{} &{} \\[-1.5ex]
1& peculiar velocities: relative rms &  20~kpc$\lap r\lap 1$~Mpc & 			$0.20e^{\pm 0.4}$ \\[1ex]
2 & \qquad redshift space anisotropy & 
10~Mpc$\lap r\lap 30$~Mpc & $0.30\pm 0.08$ \\[1ex]
3 & \qquad mean relative velocities & 10~Mpc$\lap r\lap 30$~Mpc & 		$0.30^{+0.17}_{-0.07}$\\[1ex]
4 & \qquad numerical action solutions & $r\sim 1$~Mpc   &  $0.15\pm 0.08$\\[1ex]
5 & \qquad virgocentric flow  & $r\sim 20$~Mpc &
	 $0.20^{+0.22}_{-0.15}$\\[1ex]
6 & weak lensing: galaxy-mass & 100~kpc$\lap r\lap$ 1~Mpc &
	$0.20 ^{+0.06}_{-0.05}$ \\[1ex]
7 & \qquad mass-mass & 300~kpc$\lap r\lap  3$~Mpc &
	 $0.31\pm 0.08$\\[1ex]
8 & angular size distance $r(z)$: SNe & $r\sim 3000$~Mpc &
	$0.29 ^{+0.05}_{-0.03}$\\[1ex]
9 & \qquad cluster baryon fraction  & $r\sim 3000$~Mpc & $\lap 0.3$\\[1ex]
10 & local cluster baryon mass fraction & $r\sim 10$~Mpc &
	$0.27\pm 0.02$\\[1ex]
11 & cluster mass function &$r\sim 10$~Mpc & $0.17\pm 0.05$\\[1ex]
12 & mass fluctuation power spectrum &$r\sim 100$~Mpc &
	$0.23 \pm 0.02$\\[1ex]
13 & integrated Sachs-Wolfe effect & $r\sim 300$~Mpc &
	$\sim 0.3$\\[1ex]
\hline
\end{tabular}\label{tab1} }
\vspace*{-13pt}
\end{table}

\section{The Mean Mass Density}

Table~1 lists measurements of the large-scale mean mass density. Many analyses nowadays employ joint fits to several of the observations listed in the table. That is efficient if we are working with the correct theory, but for the purpose of exploring what aspects of the theory are tested by the measurements it is best to separate the results from each physical effect or each significantly different means of observing an effect. 

My comments, with sample references, may be too brief for those who do not follow this subject and redundant for those who do, but I hope the point is clear:  quite different aspects of the theory and observations enter the thirteen measurements of nominally the same quantity, and the degree of consistency of the results checks the considerable variety of potential sources of systematic errors in the theory and observations.

\subsection{Mass Estimates}

The third column in the table indicates a characteristic range of length scales to which each measurement is sensitive. Here and throughout I adopt the distance scale represented by Hubble's constant 
\beq H_o=70\hbox{ km s}^{-1}\hbox{ Mpc}^{-1}.\label{eq:Ho}\eeq
The estimates of the mass density parameter $\Omega _m$ --- the ratio of the mean mass density to the critical Einstein-de Sitter value --- in the last column scale with the value of $H_o$ in a variety of ways, but the effect of the uncertainty in $H_o$, perhaps 20~percent, on the scatter among results is smaller than the measurement errors. 

The first seven estimates in Table~1 assume optically selected galaxies are useful mass tracers (where I mean by ``useful'' that  the error introduced by this assumption is not much larger than the measurement error). I review the case for this assumption after commenting on these seven results. 

The first two entries are based on the galaxy two-point correlation function in redshift space, where the distance of a galaxy with redshift $z$ is $cz/H_o$. At separations $r\lap 1$~Mpc the random motions of galaxies elongate the clustering in redshift space along the line of sight. I have used an old measurement\cite{DP} of this small-scale effect in entry~1 because more recent surveys tend not to probe small scales. On larger scales the redshift space correlation function is flattened along the line of sight by the streaming motion accompanying the growth of clustering. Modern surveys beautifully demonstrate this effect.\cite{Hawkins} The third entry probes the streaming by the measurement of the mean relative line-of-sight velocity as a function of separation.\cite{RomanJ}  The next entry grew out of the remark by Kahn and Woltjer\cite{LG} that the two large members of the Local Group of galaxies appear to have mass considerably larger than the observed star mass, in the approximation of a two-body system. Entry 4 comes from a more elaborate solution\cite{action}  that takes account of the mass distribution out to 40~Mpc distance, but I have entered a length scale characteristic of the Local Group because the motions of the galaxies in and near this system are the important aspect of the solution. The next entry is derived from the density contrast\cite{DP2} and streaming flow\cite{VCflow} of galaxies in the Local Supercluster, which is centered on the Virgo Cluster, at about 20~Mpc distance, with us near the edge.

 Entries 6 and 7 are based on weak gravitational lensing. The first uses the mean shear in the images of distant galaxies caused by the gravitational deflection by the masses in nearer galaxies, in the shear-galaxy cross correlation function measured by McKay {\it et al.}\cite{McKay} and analyzed in Ref.~\refcite{FP}. In entry~7 I quote the measurement and analysis of the shear autocorrelation function of galaxy images at redshifts $z\sim 1$ by Rhodes {\it et al}.\cite{Rhodes}  

The assumption that galaxies are useful mass tracers is tested by the consistency of the estimates of $\Omega _m$ from a considerable range of lengths, $100\hbox{ kpc}\lap r\lap 30\hbox{ Mpc}$, because if galaxies were biased mass tracers one might have thought the bias would vary with the length scale. An elegant and more direct test compares the galaxy three-point correlation function, scaled by the square of the two-point function, to the same expression for the mass correlation functions computed in lowest nontrivial order in perturbation theory.\cite{Fry} The latter assumes Gaussian initial conditions and models the relation between the galaxy and mass density contrasts as
\beq  
\delta _{\rm g} = b_1\delta _\rho + b_2\delta _\rho ^2/2.
\label{eq:biaseq}
\eeq
Feldman {\it et al.}\cite{FeldmanFry} give a striking illustration of the consistency of functional forms of the galaxy and mass functions in this model. Their fit to the clustering of galaxies selected by the IRAS satellite detection at $\lambda\sim 100\mu$ requires constants $b_1\sim 0.8$ and $b_2\sim  -0.4 b_1^2$, consistent with the picture that IRAS galaxies are more smoothly distributed than the mass and particularly avoid the densest regions. This agrees with the observation that gas-rich galaxies that can be luminous in the far infrared avoid dense regions where interactions more readily deplete interstellar gas. For optically selected galaxies $b_1$ is quite close to unity and $b_2$ is close to zero,\cite{Verde} consistent with the picture that these galaxies trace mass on scales $\sim 5$ to 30~Mpc. This test assumes Gaussian initial conditions, but if the result were an artifact of this assumption the agreement of functional forms would be a curious coincidence. The lessons are that galaxies need not trace mass, as illustrated by the IRAS sample, but that Nature seems to have been kind enough to have made optically selected galaxies useful mass tracers. 

Entry 8 refers to measures of the angular size distance $H_oa_or(z)$ as a function of redshift from the redshift-magnitude relation for Type Ia supernovae. I quote the fit\cite{Riess} for zero space curvature. A similar result is obtained when the supernovae data are supplemented by $r(z)$ estimates for radio galaxies.\cite{Daly} Since this describes the general expansion of the observable universe I have entered the present Hubble length as a measure of the characteristic length scale.

The next two entries are based on the idea that rich clusters of galaxies likely are large enough to have gravitationally collected a close to fair sample of baryons and dark matter. The constraint on $r(z)$ from the condition that the cluster baryon mass fraction is independent of redshift indicates that $\Omega _m$ is close to zero if $\Lambda =0$ and is about at the upper bound in Eq.~(\ref{eq:omegam}) if space sections are flat.\cite{Allenpc} Entry 10 is from the comparison of the mass fraction in clusters\cite{Allen} to the mean baryon density\cite{FP} from the standard model for the light elements; it is not sensitive to $\Lambda$. 

Entry 11 refers to the evolution of the number density of massive clusters. There are in effect two constraints: the number densities of clusters at low redshift and high. A fit\cite{Bahcall} to the two free parameters offered by the standard cosmology with negligible space curvature yields the value of $\Omega _m$ in the table and an amplitude of the primeval scale-invariant mass density fluctuations that agrees with the fluctuations in galaxy counts, yet another indication that galaxies are useful mass tracers. The length scale here and in entry 10 is the radius from which the cluster mass is gathered.

Part 3b in the standard cosmology fixes in linear perturbation theory the functional form of the mass autocorrelation function $\xi (r)$ --- for example, $\xi (r)$ passes through zero at $r = 30\Omega _m^{-1}$~Mpc (for $H_o$ in Eq.~[\ref{eq:Ho}]) --- so if galaxies trace mass the shape of the large-scale galaxy two-point correlation function offers a measure of $\Omega _m$. Entry 12 is from the  2dF galaxy redshift survey.\cite{Percival}

If space curvature or $\Lambda$ make an appreciable contribution to the present expansion rate then in the standard model the 3~K CBR radiation is perturbed at low redshift by the evolving gravitational potential of the mass distribution. Boughn and Crittenden\cite{Boughn} show that the effect is detected with reasonable confidence in the cross correlation of the WMAP CBR anisotropy measurement\cite{Bennett} with the angular distributions of radio and X-ray sources at redshifts close to unity. The constraint is not tight ---  the effect is about what is expected if $\Omega _m\sim 0.3$ and space sections are flat --- but it is an elegant convergence of theory and observation. 

I have not entered the comparison of the expansion time from high redshift with stellar evolution and radioactive decay ages because the constraint on $\Omega _m$ is somewhat looser (as shown in Table~15 in Ref.~\refcite{Freedman}), but it  certainly is an important test of the cosmology. 

\begin{figure}[ht]
\centerline{\epsfxsize=3in\epsfbox{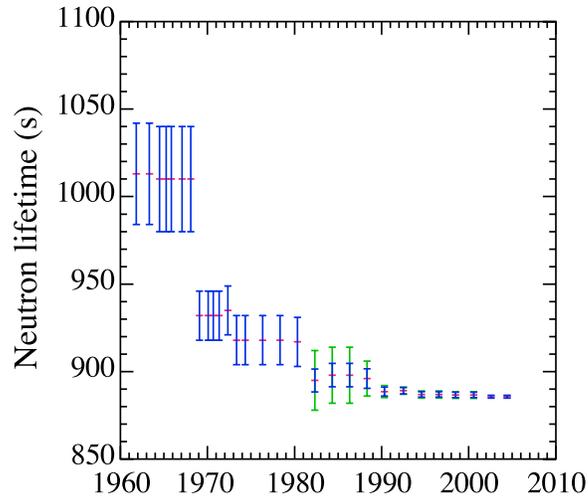}}   
\caption{Evolution of the handbook value of the neutron mean life.\label{figure}}
\end{figure}

\subsection{Systematic Errors}

The considerable number of entries in Table~1 allows us the luxury of considering the considerable variety of potential sources of systematic error, under the headers of astronomy, physics and sociology. 

To be confident about our deductions of aspects of fundamental physical reality from limited observations of processes operating on the far side of the visible universe we need to demonstrate  consistency of evidence from a broad variety of astronomical methods that are subject to different uncertainties. I conclude that for the purpose of constraining the value of  $\Omega _m$ at the modest precision of the entries in Table~1 this condition is satisfied: we can trust the astronomy. 

We must consider that we have adjusted the theory to fit the observations: that is how we arrived at part 3b of the standard model. The case that this is more than curve fitting is supported by the magnificently successful fit of the standard model to the WMAP measurements of the CBR anisotropy spectrum, with parameters that agree with independent estimates of the baryon mass density and the slope and normalization of the mass fluctuation power spectrum.\cite{Bennett} We cannot know for sure that there is no viable modified --- or alternative --- model for structure formation, however, so it is important that most of the entries in Table~1 are not sensitive to the details in part 3b, and that the consistency with entries 11 and 12, which do depend on this part of the standard model, adds to the evidence that the $\Lambda$CDM cosmology is a useful approximation. 

We also have to bear in mind that scrupulously careful measurements can be influenced by our respect for social norms. The reluctance to stray more than one or two standard deviations from what is generally accepted is illustrated by variations with time in the handbook values of physical constants, as in the example\cite{PDG} in figure~\ref{figure}. One sees a similar effect in the history of estimates of $\Omega _m$. In 1980, using the method in entry 1 in Table~1, I got\cite{Peeb} $\Omega _m = 0.65\pm 0.25$. My prejudice then was no secret: I argued that the only reasonable case is the Einstein-de Sitter model, with $\Lambda =0$ and $\Omega _m=1$, and I was glad to see that my estimate was not significantly off the right answer. A few years later the inflation concept led a good part of the community to accept that the Einstein-de Sitter model very likely is the right answer. You can find in the literature of the following decade  careful observational analyses that indicate consistency with Einstein-de Sitter. Different observational advances drove different groups away from this consensus. It happened to me in 1982, with the demonstration\cite{DP} of small relative velocity dispersion in the field (from the Center for Astrophysics redshift catalog), together with the absence of the segregation of normal and dwarf galaxies that I would have expected to see as a consequence of the biasing picture that was introduced to save this phenomenon.  By the early 1990s the large scales of clustering of galaxies and clusters of galaxies\cite{Blumenthal}$^,$\cite{Efstathiou}$^,$\cite{bahcallcen} 
and the cluster mass function\cite{bahcallcen} and baryon mass fraction\cite{White} had added to the evidence for low $\Omega _m$. But the general --- and unnervingly abrupt --- swing of opinion away from the Einstein-de Sitter model was triggered in 1998 by the supernovae redshift-magnitude measurements.\cite{Riess1}$^,$\cite{Perlmutter} 

It takes nothing away from the enormous amount of careful work by many people to remark that that social pressure likely has narrowed the scatter in Table~1. Could the apparent consistency of evidence summarized in this table be a social construction, like the case for the Einstein-de Sitter model ten years ago? Each of us may decide whether there are enough checks to rule out significant systematic errors from this source as well as the astronomy and physics. The much broader variety of evidence developed since the 1990s, and the larger number of people who are probing for something new and interesting to do with the evidence, lead me to conclude that we have reached a compelling case for real consistency at the accuracy indicated by the general run of error flags in Table~1. 

\section{The Cosmological Constant}

Is the $\Lambda$ term in the standard cosmology a social construction? The accompanying evidence for flat space sections
was welcomed as a prediction of inflation, but you can count that welcome as an effect of social pressure because inflation is a social construction, which is to say that it is a promising working hypothesis that awaits searching scientific tests. It is important that there are independent lines of evidence for the detection of $\Lambda$, mainly from measures of the angular size distance as a function of redshift,\cite{Riess} and from the WMAP measurement of the CBR anisotropy.\cite{Bennett}  The latter is somewhat beclouded by its dependence on a structure formation model with anomalies that, if real, will drive adjustments of the model and maybe of the constraint on $\Lambda$, and it is beclouded also by the puzzle of the quantum vacuum energy density, which might drive adjustments of the world picture\cite{Vilenkin} or of the gravity theory\cite{Adelberger}$^,$\cite{Dvali} and the interpretation of the cosmological tests. One can make a similar list of hazards for each estimate of $\Omega _m$, of course; the big difference is in the lengths of the lists of independent evidence. The issue of Einstein's cosmological constant has been under discussion since 1917. I suggest we wait a few more years to see how the evidence develops --- as from the cluster baryon mass fraction,\cite{Allen} the cluster mass function,\cite{Bahcall} and the expansion time scale\cite{Freedman} --- before making a definite decision about the reality of this curious term. 

\section{Testing Gravity Physics on the Scales of Cosmology}

In the 1960s, during the development of many of the precision tests of gravity physics on the scale of the Solar System and smaller, the PPN formalism offered a helpful way to judge how well the tests are doing: find the ranges of parameters allowed by the measurements, and see whether the allowed ranges include general relativity. We do not have an analogous formalism suitable for cosmology. The purpose of these qualitative remarks is to show why it is time to develop one. 

The first seven entries in Table~1 make direct use of the inverse square law for gravitational acceleration and the assumption that galaxies are useful mass tracers. It would be absurd to imagine that a failure of both assumptions conspired to produce the apparent consistency of measurements over the range of scales from 100~kpc to 30~Mpc: we may conclude that the inverse square law passes this test, within the still substantial error budget. 

The inverse square law has the special property that it preserves the functional form of the mass distribution in the growing mode of the departure from homogeneity in linear perturbation theory (because in this approximation the time derivative of the mass density contrast $\delta\rho /\rho$ is proportional to the divergence of the peculiar velocity field, which is proportional to the peculiar gravitational acceleration $\vec g$, and the inverse square law says the divergence of $\vec g$ is proportional to $\delta\rho /\rho$). A failure of the inverse square law would cause the rate of growth of the Fourier amplitudes of the mass distribution to  depend on the wavelength.\cite{Peeb1}$^,$\cite{Jimenez} This means we have another check of the inverse square law from the apparent consistency of stories for the growth of clusters of galaxies, the shape of the large-scale mass fluctuation spectrum, the CBR anisotropy spectrum, and the detection of the contribution to the anisotropy by the ISW effect. 

Entries 6 and 7 take account of the factor of two difference between the relativistic gravitational deflection of light and the naive Newtonian model. We know the factor of two is present on the scale of the Solar System. We have good reason to expect the factor of two is present on the scales of extragalactic astronomy, but I hope it is agreed that an actual observational check would be deeply satisfying. 

Under the assumptions of standard local physics and a metric description of spacetime the angular size distance $r(z)$ fixes the redshift-magnitude relation, the expansion parameter $a(t)$ fixes the expansion time as a function of redshift, and both are used in the predictions of counts of conserved objects as a function of redshift and of the anisotropy of the CBR produced by the physics of decoupling. In the standard cosmology these two functions are determined by the Friedmann equations,
\beq
\left(\dot a\over a\right)^2 = {8\over 3}\pi G\rho + {1\over a^2R^2} + {1\over 3}\Lambda ,\qquad
{\ddot a\over a} = -{4\over 3}\pi G(\rho + 3p) + {1\over 3}\Lambda ,
\label{eq:FLeq1} 
\eeq
where $aR$ is the radius of curvature of space sections in the Robertson-Walker line element. The second expression --- from the time derivative of the first --- illustrates the contribution of the pressure to the active gravitational mass density. The angular size distance may be represented by  the Sachs equation for the proper area $A={\rm d}^2$ of a beam of light running from a distant object to an observer at ${\rm d}=0$. In an exactly homogeneous and isotropic Friedmann-Lema\^\i tre model the Sachs equation is
\beq
{d^2{\rm d}\over dt^2} - {\dot a\over a}{d{\rm d}\over dt} = 
- 4 \pi G (\rho + p){\rm d}. \label{eq:sachs}
\eeq
The cosmological constant $\Lambda$ cancels here, but in models with evolving dark energy density, where $p_\Lambda\not=-\rho_\Lambda$, there is a contribution from the dark energy. The role of pressure in these examples, and in the dynamics of relativistic fluids, is elegant, and it is checked by the standard models for the light elements and the CBR anisotropy. A formal analysis of this check would be elegant too. 

A theme of this essay is that we gain insight on what we are learning from the study of cosmology by considering how well we can constrain the fundamental elements of the subject. The present state of the tests draws special attention to  the empirical evidence for the gravitational inverse square law and its relativistic generalizations. The evidence may be organized in three categories by the ranges of characteristic length scales, 
\beqa
&&10^{-1}\hbox{ cm}\lap r_{\rm a}\lap 10^{13}\hbox{ cm}, \nonumber\\ 
&&10^{13}\hbox{ cm}\lap r_{\rm b}\lap 10^{21.5}\hbox{ cm}, \label{eq:scales}\\ 
&&10^{23.5}\hbox{ cm}\lap r_{\rm c}\lap 10^{27}\hbox{ cm}.
\nonumber
\eeqa
Class (a) represents the precision tests, from the laboratory and from measurements in the Solar System and binary pulsars.\cite{Adelberger}  The class (b) evidence is that the inverse square law continues to predict the gravitational acceleration, to a factor of two or so, to about 1~kpc, on the basis of the consistency of dynamical mass estimates in the inner parts of large galaxies with estimates of the seen masses in stars. In the factor $\sim 100$ gap to the class (c) results from cosmological tests something happens that drives us to the postulate of dark matter with about five times the mass of the baryons. Since the standard theories of light elements and the CBR anisotropy require this dark matter we may expect to have a convincing closure of this gap. The modest accuracy in class (c) extends to $\sim 30$~Mpc, and to reasonable evidence of a detection, in the ISW effect, at $\sim 300$~Mpc. There are even better examples of the ability of standard physics to organize phenomena over enormous ranges of scales, but Eq.~(\ref{eq:scales}) is impressive enough, and cosmology is enjoying a growth spurt that will continue to improve the tests of gravity physics on really large scales. 

\section*{Acknowledgments}

I am grateful for advice from Steve Allen, Neta Bahcall, Masataka Fukugita, Will Percival, and Bharat Ratra.

\end{document}